\documentclass[12p,preprint]{aastex}
\usepackage{graphicx}
\begin{document}
\title{Electron Surfing Acceleration in Magnetic Reconnection}
\author{Masahiro Hoshino}
\affil{Department of Earth and Planetary Science,
University of Tokyo, Bunkyo, Tokyo, JAPAN}
\begin{abstract}
We discuss that energetic electrons are generated near the X-type magnetic reconnection region due to a surfing acceleration mechanism.  In a thin plasma sheet, the polarization electric fields pointing towards the neutral sheet are induced around the boundary between the lobe and plasma sheet in association with the Hall electric current.  By using a particle-in-cell simulation, we demonstrate that the polarization electric fields are strongly enhanced in an externally driven reconnection system, and some electrons can be trapped by the electrostatic potential well of the polarization field.  During the trapping phase, the electrons can gain their energies from the convection/inductive reconnection electric fields.  We discuss that relativistic electrons with MeV energies are quickly generated in and around the X-type neutral region by utilizing the surfing acceleration.
\end{abstract}
\section{Introduction}
Magnetic reconnection is believed to play an important role on plasma heating and acceleration over a broad range of plasma phenomena.  Observations of magnetic storms/substroms in terrestrial magnetosphere and of solar and stellar flares have shown that enormous hot plasmas with temperature of $10^6$ to $10^8$ K are generated explosively through magnetic reconnection.  In these widespread observations, not only the hot thermal plasma population but also nonthermal particles of MeV energy have been reported to be quickly generated in association with the hot plasmas [e.g., {\it Baker and Stone}, 1976; {\it Lin et al.}, 2003], and the production of energetic particles has been recognized as one of the important aspects of reconnection.   So far many theoretical and observational studies on plasmas heating during magnetic reconnection have been conducted, and it is accepted as being beyond doubt that reconnection is one of essential key processes of magnetic energy conversion.  However, the understanding of the non-thermal particle acceleration mechanism remains poor.

In order to understand the mechanism of particle acceleration, many people have studied the charged particle motion under the reconnection electric and magnetic fields, and have conducted test particle simulation of the charged particle motion for the prescribed reconnection fields obtained from the non-linear MHD simulations [e.g., {\it Sato et al.}, 1982; {\it Scholer and Jamitzky}, 1987; {\it Birn and Hesse}, 1994; {\it Kliem}, 1994].  It is discussed that the high energy particle can be effectively generated under the reconnection geometry.  The acceleration is mainly provided by the direct acceleration of particle with the $v \times B$ reconnection electric field.  It is also suggested that the magnetohydrodynamic turbulence can contribute to the particle energization as well [e.g., {\it Ambrosiano et al.}, 1988].  Recently {\it Hoshino et al.} [2001a; for convenience we shall call this paper as Paper 1] studied the suprathermal electron acceleration in a particle-in-cell (PIC) simulation where both electromagnetic fields and motion of charged particles are self-consistently solved.  They found that electron acceleration first occurs near the X-type region due to the meandering/Speiser motion [{\it Speiser}, 1965], and those pre-accelerated electrons are further accelerated by $\nabla$ B and/or curvature B drift in the magnetic field pile-up region produced by compression of the reconnection outflow.  They demonstrated that the spurathermal electrons of several 10 keV are generated during reconnection in magnetotail, but the acceleration efficiency was not necessarily strong for producing MeV electrons.

Non-adiabatic motion of particle is of major importance in energetic particle acceleration.  The particle scattering by turbulent electromagnetic waves is recognized as one of the essential processes of breaking down the adiabatic motion of particle, and it is the basis of the efficient energy transfer process between the particle and the waves.  In fact, this kind of scattering process is confirmed in the numerical simulation in Paper 1, but it was not a main player of particle acceleration in their two-dimensional PIC simulation.  Another class of non-adiabatic process can be realized when the gyro-radius of the particle becomes comparable to a characteristic scale length of the magnetic field curvature [{\it B\"{u}chner and Zelenyi}, 1989; {\it Delcourt et al.}, 1996].  In this regime, the particle is known to show a chaotic behavior, and the high energy particles can be efficiently accelerated by the interaction of the complex particle trajectory under the reconnection electric field.  In the numerical simulation, this 2nd class of non-adiabatic motion was observed in the pile-up region of the reconnecting magnetic field lines where the reconnection outflow collides with the pre-existing plasma sheet plasma.

It is important to mention that many thin current sheets are often observed during the plasma sheet crossing of satellites in the terrestrial magnetotail [{\it Sergeev et al.}, 1993; {\it Pulkkinen et al.}, 1994; {\it Mukai et al.}, 1998], and the thickness of the current sheet is found to be of order of ion inertia length [{\it Fairfield, 1984}; {\it McComas et al.}, 1986; {\it Mitcheel et al.}, 1990; {\it Asano et al.}, 2003].  Moreover, during the reconnection events with fast plasma flows, a quadrupole magnetic field structure generated by the Hall electric current and non-gyrotropic ions distributions have been reported in the magnetotail observation [e.g., {\it Hoshino et al.}, 1998; {\it Nagai et al.}, 2001].  These observations suggest that the thickness of the current sheet becomes as thin as the ion inertia length just before and during reconnection in magnetotail.
  
In such a thin current sheet, energy conversion through kinetic plasma processes plays an important role on plasma sheet dynamics and evolution.  The micro-scale reconnection structure has made remarkable advance lately in both theory/simulation and observational studies [e.g., {\it Nishida}, 2000; {\it Birn et al.}, 2001].  Thanks to the modern supercomputers, it is now possible to discuss the electron dynamics embedded in a macro-scale reconnection structure [e.g.,{\it  Hesse et al.}, 2001; {\it Shinohara et al.}, 2001; {\it Lapenta et al.}, 2002; {\it Scholer et al., 2004}].  {\it Hoshino et al.} [2001b] demonstrated that the strong Langmuiur waves are generated in association with a bump-on tail electron distribution function in the plasma sheet boundary.  {\it Drake et al.} [2003] performed the magnetic reconnection simulation with a guide magnetic field parallel to the electric current in the plasma sheet, and they observed the intense bipolar electric fields and electron heating near the magnetic X-line and separatrices through an electron beam instability.  

Although those intense high-frequency waves observed in the large-scale reconnection simulations may contribute to the particle acceleration through the wave scattering and/or the direct acceleration along the magnetic field, we would like to propose another more efficient acceleration process in this paper.  We focus on non-thermal electron acceleration by the ambipolar electric fields generated around the plasma sheet boundary.  Firstly we demonstrate by using the PIC simulation that intense ambipolar electric fields are generated under a driven reconnection, namely, in the presence of an external driving plasma flow [e.g., {\it Horiuchi and Sato}, 1999].  Then we discus that the localized ambipolar electric field is particularly important for supra-thermal electron acceleration through the so-called surfing acceleration [e.g., {\it Sagdeev and Shapiro}, 1973; {\it Katsouleas and Dawson}, 1983].  We argue that relativistic electrons with MeV energies can be quickly generated by virtue of the surfing acceleration.

%%%%%%%%%%%%%%%%%%%%%%%%%%%%%%%%%%%%%%%%%%%%%%%%%%%%%%%%%%%%%%%%%%%%%%%%%%%%
\section{Strong polarization electric fields by driven reconnection}
We study the nonlinear evolution of the collisionless driven magnetic reconnection by using the two-dimensional, particle-in-cell simulation code.  The simulation parameters are basically same as those discussed in Paper 1, but we continuously drive a magnetized plasma flow from both the top and bottom lobe regions so as to initiate a fast reconnection [e.g., {\it Horiuchi and Sato}, 1999; {\it Birn et al.}, 2005].  The physical system size is $L_x/\lambda = 38.4$ and $L_z/\lambda = 32$, where $\lambda$ is the thickness of the initial plasma sheet.  We assumed the double periodic boundary in the x and z directions, but we set the plasma injection regions at $L_z/\lambda=\pm 12$, where the magnetized plasma consisting of electrons and ions is injected.  The external driven electric field at the injection boundary are given by $E_{dri}/E_0 = \alpha_1/{\rm cosh}^2(x/L_{dri})+\alpha_2$ where $L_{dri}/\lambda = 2$ and $E_0 = V_A B_0/c$.  Two cases are simulated: one is the weakly driven case with $\alpha_1 = \alpha_2 =0.1$, and the other is the strongly driven case with $\alpha_1 = \alpha_2 =0.2$.  The reconnection structures in the nonlinear saturation stages are almost same, but for the weakly driven case it takes twice as long as the nonlinear evolution for the strongly driven reconnection.  In this paper, we discuss the result for the strongly driven case.

The numerical grid points are $N_x \times N_z = 1536 \times 1280$, and the number of total particles is about $4 \times 10^8$.  The plasma parameters are as follows: Ion inertia length $V_A/\Omega_i$ is 1.15 $\lambda$ with 46 grids, and electron inertial length $c/\omega_{\rm pe}$ is 0.14 $\lambda$ with 6 grids.  The mass ratio of ions to electrons $m_i/m_e$ is 64, for simplicity.  The ratio of the speed of light to the Alfven velocity $c/V_A$ is 16.8.  The plasma temperatures are set as follows: $T_{\rm i,ps}/T_{\rm e,ps}=4$, $T_{\rm i,lobe}/T_{\rm i,ps}=1/25$, and $T_{\rm i,lobe}=T_{\rm e,lobe}$, where the subscript ``i" means ion, ``e" electron, ``ps" the plasma sheet, and ``lobe" the lobe, respectively.  The ratio of the Alfven velocity $V_A$ to the ion thermal velocity $v_{th,i}=\sqrt{2 T_{\rm i,ps}/m_i}$ is 1.15, where the Alfven velocity is defined by the lobe magnetic field $B_0$ and the plasma sheet density $n_0$.  The initial lobe plasma density is set to be 5 $\%$ of the plasma sheet density $n_0$.

Figure 1 shows a snapshot of the nonlinear stage of a driven reconnection at $t /\tau_A = 52.2$, where $\tau_A = \lambda/V_A$ is the Alfven transit time.  In the top panel, the color contour of the electric field $E_x$ is shown together with the magnetic field lines denoted by the solid lines, while the bottom panel shows the electric field $E_z$.  We find that a pair of V-shaped strong electric fields are induced almost along the plasma sheet boundary, namely around the boundary between the lobe and the plasma sheet.  The electric fields point towards the plasma sheet, which directions are almost perpendicular to the local magnetic field line.  These ambipolar electric fields are known to be produced in association with Hall electric current in a thin current sheet due to the inertia difference between ions and electrons, when the thickness of the plasma sheet becomes comparable or less than the ion inertia length [{\it Hoh}, 1966; {\it Hoshino}, 1987].  For the thin plasma sheet, ions are unmagnetized and do not move with the magnetic field line motion, while electrons remain magnetized and are frozen in the magnetic field line.  Therefore, the electron density becomes higher than the ion density in the thin plasma sheet, and an ambipolar/polarization electric field is produced in order to maintain the charge neutral plasma.  In non-linear stage of reconnection, those electric fields are localized in the boundary region.  From our simulation results, we find that the amplitude of the polarization electric fields reaches up to about 10 -- 20 times of the driven electric field $E_{dri}$, and the typical region width where the polarization electric fields are induced is less than the ion inertia scale.

In addition to the above large scale polarization electric field, we also find a series of large amplitude electrostatic wave emission propagating outward from the X-type region in the boundary region.  In the spontaneous reconnection in our previous study [{\it Hoshino et al.}, 2001b], we found a signature of the Langmuir wave emission, but the amplitude remained too small to recognize it as a series of wave trains.  The large amplitude waves in Figure 1 can be produced as the result of the nonlinear evolution of the Langmuir waves in the driven reconnection.  It is interesting to note that {\it Drake et al.} [2003] reported the similar bi-polar electric field generation near the X-type region with a guide magnetic field parallel to the electric current, namely, the situation for the magnetopause reconnection.  

In the bottom panel of Figure 1, we find that $E_z$ is enhanced for both top and bottom regions around $(X, Z) =(0, \pm 4)$.  The top region has the positive $E_z$, while the bottom is the negative one.  These regions with a finite electric field $E_z$ correspond to the front side of the compressed plasma propagating from the driven boundary at $Z/\lambda = \pm 12$.  We would like to mention that the time evolution of reconnection was speeded up when the location of the driven boundary is set to be closer to the neutral sheet, but in order to get the better separation between the X-type reconnection structure and the above front side region of the driven flow, we chose the boundary location of the external magnetized plasma flow at $L_z/\lambda=\pm 12$.

It is also noteworthy that the external driven system dramatically changes the time evolution of system [e.g., {\it Hayashi and Sato}, 1978; {\it Hoshino}, 1991; {\it Pritchett and Coroniti}, 1995; {\it Horiuchi and Sato}, 1999; {\it Birn et al., 2005}].  In the driven current sheet, the current layer becomes thin with increasing the lobe magnetic field.  The time evolution of the electric current at the neutral sheet $J_{ns}$ for one-dimensional slab geometry can be expressed by [{\it Drake et al.}, 1981; {\it Hoshino}, 2004]
\begin{equation}
  J_{ns}(t) \approx
J_{ns}(0) \left( \frac{B_{dri}(t)}{B_{dri}(0)} \right)^{4/\Gamma},
\end{equation}
where $B_{dri}$ is the amplitude of magnetic field at the driven boundary, and $\Gamma$ is the ratio of specific heat.  Then the time evolution of thickness of the driven current sheet $\lambda_{dri} \propto B_{dri}/J_{ns}$ can be evaluated as,
\begin{equation}
  \frac{\lambda_{dri}(t)}{\lambda_0} \approx 
       \left( \frac{B_{dri}(t)}{B_{dri}(0)} \right)^{1-4/\Gamma},
\end{equation}
where $\lambda_0$ is the initial current sheet thickness.  For the case of one-dimensional plasma sheet thinning, $\Gamma$ may be taken to be $2$ because of two-dimensional velocity space compression.  However, if the time scale of the isotropization of the anisotropic velocity distribution function is faster than the plasma compression time scale, we can use $\Gamma = 5/3$.  If we could assume an isothermal state with $\Gamma = 1$, $\lambda_{dri} \propto B_{dri}^{-3}$, the thickness of the current sheet is crucially affected by the boundary magnetic field.  The reason why the strong influence appears for an isothermal state is that the thermal energy is extracted due to the heat flux escape from the neutral sheet, and that in order to maintain the pressure balance the current sheet quickly shrinks.  For the case of non-uniform external driven boundary condition such as $E_{dri} \propto {\rm cosh}^{-2}(x)$, it is easily expected that the heated plasma at the center of the highest density region can be removed along the magnetic field line.  Having those effects in our mind, we may expect that the driven force in the reconnection setting has a strong influence on the current sheet thinning even before the onset of reconnection. 

%%%%%%%%%%%%%%%%%%%%%%%%%%%%%%%%%%%%%%%%%%%%%%%%%%%%%%%%%%%%%%%%%%%%%%%%%%%%
\section{Nonthermal electrons and surfing acceleration}
Shown in Figure 2 is energy spectra of electrons integrated over pitch angle in the whole reconnection region.  The energy $\varepsilon$ is defined as $\varepsilon = (\gamma -1) m_ec^2$, where $\gamma$ is the Lorentz factor of the particle.  Two dashed curves show the spectra at an early stage of reconnection at $t/\tau_A=7.5$ and the nonlinear stage at $t/\tau_A=44.7$ before the onset of the strong V-shaped polarization electric field seen in Figure 1.  The spectra with the solid curves are about the well-developed nonlinear stages at $t/\tau_A=52.2$ and $t/\tau_A=59.6$ when the electrostatic wave emission becomes strongly active.  As the time goes on, one can find the relativistic electrons are generated, and a large amount of relativistic electrons are quickly produced, and the nonthermal energy density normalized by the thermal energy one becomes $10 \%$.

In the spontaneous reconnection discussed in Paper 1, we already found the production of suprathermal electrons, but the nonthermal electron energy density was about $1\%$ compared with the thermal energy density.  We will discuss later that the difference of energization is due to the amplitude of a pair of the V-shaped polarization electric fields excited around the boundary between the lobe and the plasma sheet.  In fact, the amplitude of the polarization electric field in the spontaneous reconnection is about $E/E_0 \sim 1$, while that in the driven reconnection is about $E/E_0 \sim 5-10$.

In order to understand how and where the high energy electrons are generated, we investigate an electron trajectory.  Shown in Figure 3 is a typical trajectory of electron that is strongly accelerated during the interaction with the plasma sheet.  The solid curve is the trajectory in $(X, Z)$ space, and the dashed curve shows the energy history as a function of $X$.  The trajectory starts at $(X,Z) \simeq (8.7, 2.3)$ denoted by the position $1$, where the electron is situated in the lobe region, and the electron remains a very low energy.  As the electron is moving toward the X-type neutral line, it gradually gains the energy of $\varepsilon/m_e c^2 \sim 0.3$ at $(X,Z) \simeq (1, 0)$ indicated by the position $2$.  During the Speiser motion around the X-type neutral line, the electron moves from the position $2$ to $3$, and the electron gains further energy up to $ \varepsilon /m_e c^2 \sim 0.6$.  After the electron is ejected outward from the X-type neutral region, it travels toward the magnetic field line pile-up region, which may correspond to the region between the position $4$ and $5$.  In this pile-up region, $\nabla$ B/curvature B drift motion plays an important role on particle energization, and we find that the electron is accelerated up to $\varepsilon/m_e c^2 \sim 0.7$ during a chaotic motion around the position $5$.

We discussed two-step acceleration mechanism of electron in Paper 1: one is the standard Speiser/meandering motion around the X-type region, and the other is $\nabla$ B/curvature B drift acceleration for non-adiabatic particles with $\kappa \sim 1$ in the magnetic field line pile-up region, where $\kappa^2$ is the ratio of the curvature for the magnetic field and the Larmor radius [{\it B\"{u}chner and Zelenyi}, 1989].  In the driven reconnection system, the reconnection fields grow up much faster than the spontaneous reconnection, and the point of this paper is that we found the pre-stage acceleration before the Speiser acceleration.  The acceleration seems to be switched-on during the nonlinear evolution of the strong electric fields shown in Figure 1.  

For convenience we shall call the pre-acceleration as the 1st stage acceleration, the Speiser acceleration as the 2nd stage, and the acceleration in the pile-up magnetic field region as the 3rd stage acceleration.  We think that all three acceleration phases are important for the acceleration of electrons to high energies.  Figure 4 summarizes the above acceleration processes.  Except for the pre-acceleration of electron during the cold electron transport stage from the lobe toward the X-type neutral line, the acceleration processes listed here are same as those discussed in Paper 1.  

Let us discuss the 1st stage acceleration in detail.  We think that the acceleration mechanism in the 1st stage is understood by a ``surfing acceleration" due to the polarization electric field induced in the boundary between the lobe and the plasma sheet, because the acceleration occurs in the direction perpendicular to both the magnetic field and the polarization electric field.  Let us explain what the surfing acceleration is and how the surfing acceleration works in the reconnection region.  Shown in Figure 5 is a cartoon of the key elements of the electromagnetic fields in the surfing acceleration.  The dashed lines show the magnetic field lines around the reconnection region, and we take the tilted coordinate system of $(X',Z')$ where the $X'$ axis is parallel to the local magnetic field.  The polarization electric fields points the negative $Z'$ direction almost perpendicular to the local magnetic field.  The equation of motion for an electron in this coordinate system can be written as
\begin{eqnarray}
 \frac{dp_{x'}}{dt} & = & -e E_{x'} \simeq 0, \\
 \frac{dp_{y}}{dt} & = & -e E_{y} - \frac{e}{c} v_{z'} B_{x'} 
       \simeq -e E_{y}, \label{Eq_Motion_y} \\
 \frac{dp_{z'}}{dt} & = & -e E_{z'} + \frac{e}{c} v_{y} B_{x'}.
\end{eqnarray}
We have assumed that the field-aligned electric field $E_{x'}$ is negligible for the surfing acceleration, but note that the small parallel electric field may plan an important role on the thin current sheet structure [{\it Zelenyi et al.}, 2004].  In Eq.(4) we have neglected the Lorentz force term of $v_{z'} B_{x'}$, because this term remains small after averaging it during the trapping motion in the z-direction.  Let us also assume that the energy of the electron transported from the lobe is smaller than the potential energy $e \phi$, where $E_{z'} = - \partial \phi/\partial z'$.  Then the electron can be trapped by the potential well, and we assume that $v_{z'}$ is very small for the early phase.
During this trapping phase, the electron can be accelerated by the reconnection electric field $ E_y$, if the electric force of $e E_{z'}$ can balance with the Lorentz force of $e v_{y} B_{x'}/c$.  

If the Lorentz force becomes larger than the electric force during this acceleration, the particle can escape from the barrier of the polarization electric field $ E_{z'}$, and the energization of particle will stop.  The maximum velocity of the surfing acceleration can be roughly estimated by
\begin{equation}
  v_{\rm max} \sim c \frac{ E_{z'}}{ B_{x'}}.
\label{trapping_velocity}
\end{equation}
One of the key points of the surfing acceleration is that the energization of particles comes from the resonance with the $E_y$ electric field, even though the final velocity $v_{\rm max}$ is expressed by $E_{z'}$.  Another point is the localization of the $E_{z'}$ electric field.  If $E_{z'}$ were uniform in space, the surfing acceleration is replaced by the standard $E \times B$ drift.  In the $E \times B$ drift case, the energization during the initial pick-up phase can be provided from the $E_{z'}$ electric field under $ E_{z'} \gg E_y$.  

The other interesting behavior of the surfing acceleration appears under a strong $E_{z'}$ field.  If the electric field $E_{z'}$ is larger than the magnetic field $B_{x'}$, the particle cannot escape from the barrier of the potential well, and it will be able to gain quickly a relativistic energy.  This kind of acceleration does not appear for most of MHD phenomena where $E/B \sim O(V_A/c)$, but this is not the case for the driven reconnection in a thin plasma sheet, because the localized electric field can become as large as the reconnection magnetic field near an X-type region.  

Figure 6 is the velocity hodogram in $(P_y, P_z)$ for the same electron shown in Figure 3 (where $P_j = \gamma v_j/c$), and the same time interval is chosen.   The markers of positions 1 to 5 are chosen to be the same time stage as Figure 3 as well.  In the initial phase before 1st stage acceleration, the electron is located around $(P_y, P_z)=(0.0, 0.0)$.  During the 1st stage acceleration, the electron moves toward position 2 with $(P_y, P_z)=(0.8, -0.2)$ along the $P_y$ velocity axis, which suggests that the particle gains energy by traveling toward the anti-direction against the electric field $E_y$.  This behavior confirms that the DC acceleration occurs in a finite magnetic field region, which is known as the surfing acceleration [e.g., {\it Katsouleas and Dawson}, 1983; {\it Hoshino and Shimada}, 2002].  In 2nd stage of meandering/Speiser acceleration, the electron further moves to position 3 with $(P_y, P_z)=(1.0, -0.2)$, and after that the electron changes the velocity direction due to the Lorentz force and is ejected from the X-point region.  After position 5, the large circular motion shows the gyro-motion of the accelerated electron.

If the gyro-radius for the energetic particles generated during 1st and 2nd acceleration stages becomes comparable to the local curvature of the magnetic field pile-up region, the electron motion becomes non-adiabatic so that some of electrons can be further accelerated under the reconnecting magnetic fields with the mirror geometry.  From our PIC simulations, we conclude that together with the meandering/Speiser acceleration, the surfing acceleration can contribute to the production of energetic electrons whose gyro-radii are comparable to a local magnetic field curvature.  

%%%%%%%%%%%%%%%%%%%%%%%%%%%%%%%%%%%%%%%%%%%%%%%%%%%%%%%%%%%%%%%%%%%%%%%%%%%%
\section{Conclusions and Discussion}

We demonstrated for the first time the electron surfing acceleration during reconnection in a thin plasma sheet by using the PIC simulation.  However, we used a reduced mass ratio of $m_i/m_e = 64$ in order to save a computational CPU time, and it is necessary to discuss carefully the validity of the surfing acceleration in more general plasma parameters.  

One of the main points in the surfing acceleration is the amplitude of the polarization electric field in association with the Hall electric current, because the efficiency of the surfing acceleration is controlled by the amplitude of V-shaped polarization electric field induced around the X-point region.  Let us quickly estimate the amplitude of the electric field.  The electrostatic potential energy induced in the boundary may be equated to the ion thermal energy.  The ions are assumed to be in a Boltzmann-Maxwell state, because ions are unmagnetized in a thin plasma sheet, and we roughly obtain,
\begin{equation}
  e \phi \sim T_i.
\end{equation}
The thickness $l$ of the electrostatic potential $\phi$ may be estimated as an intermediate scale between the ion inertia and the electron inertia length from our simulation result, and for a moment we may assume that $l$ is estimated by the hybrid inertia scale, i.e., $l=c/\sqrt{\omega_{pe} \omega_{pi}}$.  Then the polarization electric field $E_p$ can be obtained as,
\begin{equation}
   \frac{E_p}{B_0} \sim \frac{\phi}{l B_0} 
       \sim \left(\frac{\omega_{ce}}{\omega_{pe}}\right) 
\left( \frac{c}{l \sqrt{\omega_{pe} \omega_{pi}}} \right)
\left( \frac{m_e}{m_i} \right)^{1/4},
\label{E_B_ratio}
\end{equation}
where $\omega_{ce}=eB_0/m_e c$ is defined by using the lobe magnetic field, and $\omega_{pe}$ is for the plasma sheet.  We have assumed that the ion thermal velocity is almost equal to the Alfven speed $V_A$.  The ratio of $\omega_{ce}/\omega_{pe}$ is a small value in the magnetotail.  For a typical tail parameters, 
\begin{equation}
     \frac{\omega_{ce}}{\omega_{pe}} \sim 8.8 \times 10^{-2}
         \left( \frac{B_0}{20 {\rm nT}} \right)
         \left( \frac{0.5 \rm cm^{-3}}{n_0} \right)^{1/2},
\end{equation}
and we will get the amplitude of the polarization electric field in magnetotail is 
\begin{eqnarray}
    E_p &\sim& 540 
         \left( \frac{B_0}{20 {\rm nT}} \right)^2
         \left( \frac{0.5 \rm cm^{-3}}{n_0} \right)^{1/2}
         \left( \frac{c}{\omega_{pe} l} \right) {\rm mV/m} \\
&\sim& 13 
         \left( \frac{B_0}{20 {\rm nT}} \right)^2
         \left( \frac{0.5 \rm cm^{-3}}{n_0} \right)^{1/2}
         \left( \frac{c}{\omega_{pi} l} \right) {\rm mV/m}.
\end{eqnarray}
We find that the polarization electric field ranges from 10 mV/m (for $l = c/\omega_{pi}$) to 500 mV/m (for $l = c/\omega_{pe}$).

If we normalize the porlarized electric field $E_p$ by the typical MHD electric field of $B_0 V_A /c$, we get
\begin{equation}
  \frac{E_p}{B_0 V_A /c} \sim \sqrt{\frac{m_i}{m_e}} 
           \left( \frac{c}{\omega_{pe} l} \right)
           = \left( \frac{c}{\omega_{pi} l} \right).
\end{equation}
Even if the scale length $l$ is equal to the ion inertia length of $c/\omega_{pi}$, the polarization electric field $E_p$ becomes of order of the typical MHD electric field of $E_0=B_0 V_A /c$, which is still larger than the reconnection electric field $E_*=\alpha_r E_0$ in the order of $0.1 \times B_0 V_A /c$ ($\simeq$ 1 mV/m in magnetotail), where we have assumed the reconnection rate $\alpha_r \sim 0.1$.

Next let us estimate the maximum electron energy $\varepsilon_{max}$ during the surfing acceleration.  From Eq.(\ref{Eq_Motion_y}), we obatin,
\begin{equation}
  \varepsilon_{max} \sim \int_{0}^{T_*} e E_* v_y(t) dt
  \sim \frac{(e E_*)^2}{2 m_e} T_*^2,
\end{equation}
where $T_*$ is the time interval of the surfing phase.  We have assumed that the initial electron energy is negligible compared to the final energy $\varepsilon_{max}$.  The time inverval of the surfing phase $T_*$ may be estimated as,
\begin{equation}
  T_* \sim \frac{L}{\alpha_r} \frac{1}{v_{e0}},
\end{equation}
where $L$ is the thickness of the plasma sheet where the polarization electric field is induced, and the distance along the magnetic field from the lobe to the X-type neutral point can be multiplied by the geometrical fact of $1/\alpha_r$.  $v_{e0}$ is a typical electron velocity along the plasma sheet boundary during the surfing phase, and it may have an intermediate velocity between the lobe and plasma sheet thermal velocities.  Then we obtain,
\begin{equation}
   \frac{\varepsilon_{max}}{m_e c^2} \sim \frac{\theta}{2}
       \left( \frac{\omega_{ce}}{\omega_{pe}} \right)^2
       \left( \frac{L \omega_{pi}}{c} \right)^2 
       \left( \frac{v_{th,e}}{v_{e0}} \right)
   \sim O(10^{-2}),  
\label{max_energy}
\end{equation}
where $T_i/T_e = \theta$, which is known to be about 5 in the terrestrial magnetotail.  We have assumed that $v_{th,e}/v_{e0}$ is of order of $0.1 \sim 1$.  The above equation suggests that electrons can be accelerated up to several keV during the surfing acceleration, therefore those electrons can satisfy non-adiabatic condition with $\kappa \sim 1$ for the reconnecting magnetic field line in a thin plasma sheet [{\it Hoshino et al.}, 2001a].  Thanks to this pre-acceleration during the plasma transport from the lobe toward the X-type region, $\nabla$ B and/or curvature drift as the 3rd stage of acceleration can give further energization (see Figure 4).  We think that, by combining all three acceleration steps, electrons with MeV energies can be quickly produced during magnetic reconnection. 

It should be noted that the estimation of the maximum energy in Eq.(\ref{max_energy}) is  valid only when the trapping condition of electrons with $E_{z'} \sim v_{y} B_{x'}/c$ is satisfied during the surfing acceleration at the plasma sheet boundary.  As the electron travels towards the X-type region, it is expected that the amplitude of the local magnetic field $B_{x'}$ will decrease.  If so, the trapping condition could be still satisfied for a large $v_y$, and the electron can gain energy until the maximum energy estimated by Eq.(\ref{max_energy}).  More careful analysis of the trapping condition would be necessary in order to clarify the efficiency of the surfing acceleration.

It might be interesting to note the structure of the non-thermal energy spectrum.  Shown in Figure 7 is the energy spectrum at $t/\tau_A =59.6$, which is the same energy spectrum in Figure 2 except for the double logarithmic scales.  We find that the spectrum can be approximated by a double power-law spectrum, in which the lower energy band has the spectral index of about 3 and the higher band with about 7.  The brake energy of the double power-law spectrum is about $\varepsilon_{max}/m_e c^2 \simeq 0.7$.  In our simulation parameters, we used $\theta=4$, $(\omega_{ce}/\omega_{pe})^2=0.05$, and $L\omega_{pi}/c \simeq 2$, and the typical electron velocity $v_{e0} \sim 0.5 v_{th,e}$.  Therefore, from Eq.(\ref{max_energy}), the maximum energy expected from the surfing acceleration is estimated as $\varepsilon_{max}/m_ec^2 \simeq 0.8$.  Since the maximum energy of the surfing acceleration is very close to the simulation result, we think that the lower-energy population of the double power-law energy spectrum can be generated by the surfing acceleration, and the higher-energy population is mainly produced by $\nabla$ B/curvature B drift motion.  It would be interesting to mention that the double power-law spectrum is observed in the solar corona by the RHESSI satellite [e.g., {\it Lin et al.}, 2003], and the kappa distribution functions are often observed in the terrestrial magnetotail [e.g., {\it Christon et al.}, 1989].  The surfing acceleration might be one of mechanisms to explain the observed power-law spectrum.

Finally, we should like to make the point that it may be possible to explore many aspects of the electron surfing acceleration from {\it in situ} observations in the terrestrial magnetosphere.  One of the key points of the surfing reconnection is to confirm observationally a structure of polarization electric field.  There are several observations regarding to the spiky electric field structure in the boundary between the lobe and the plasma sheet [e.g., {\it Cattell et al.}, 1994], but so far those observations are not discussed in the framework of the model presented here.  Further investigations on the small-scale electric field structure could give us better understanding on the electron acceleration in the course of magnetic reconnection.

%% ------------------------------------------------------------------------ %%
%  ACKNOWLEDGMENTS
%% ------------------------------------------------------------------------ %%

%\begin{acknowledgments}
The author especially wishes to express his thanks to J. Drake, M. Hesse, S. Imada, and P. Pritchett for valuable discussions.  He also depended on useful discussions with many participants at the Magnetic Reconnection Theory Program at the Isaac Newton Institute for Mathematical Sciences.  This was supported in part by the Institute of Space and Astronautical Science (ISAS/JAXA).  
%\end{acknowledgments}

%% ------------------------------------------------------------------------ %%
% FIGURES
%
% PLEASE NOTE: WHEN YOU SUBMIT YOUR LATEX FILE TO GEMS, COMMENT OUT ANY COMMANDS
% THAT INCLUDE GRAPHICS (see example below).
% ---------------
% ONE-COLUMN figure example
% (For further instructions see FIGURE, PLATE, AND TABLES section at end of file)

\begin{figure}
\centering
\includegraphics[width=\linewidth]{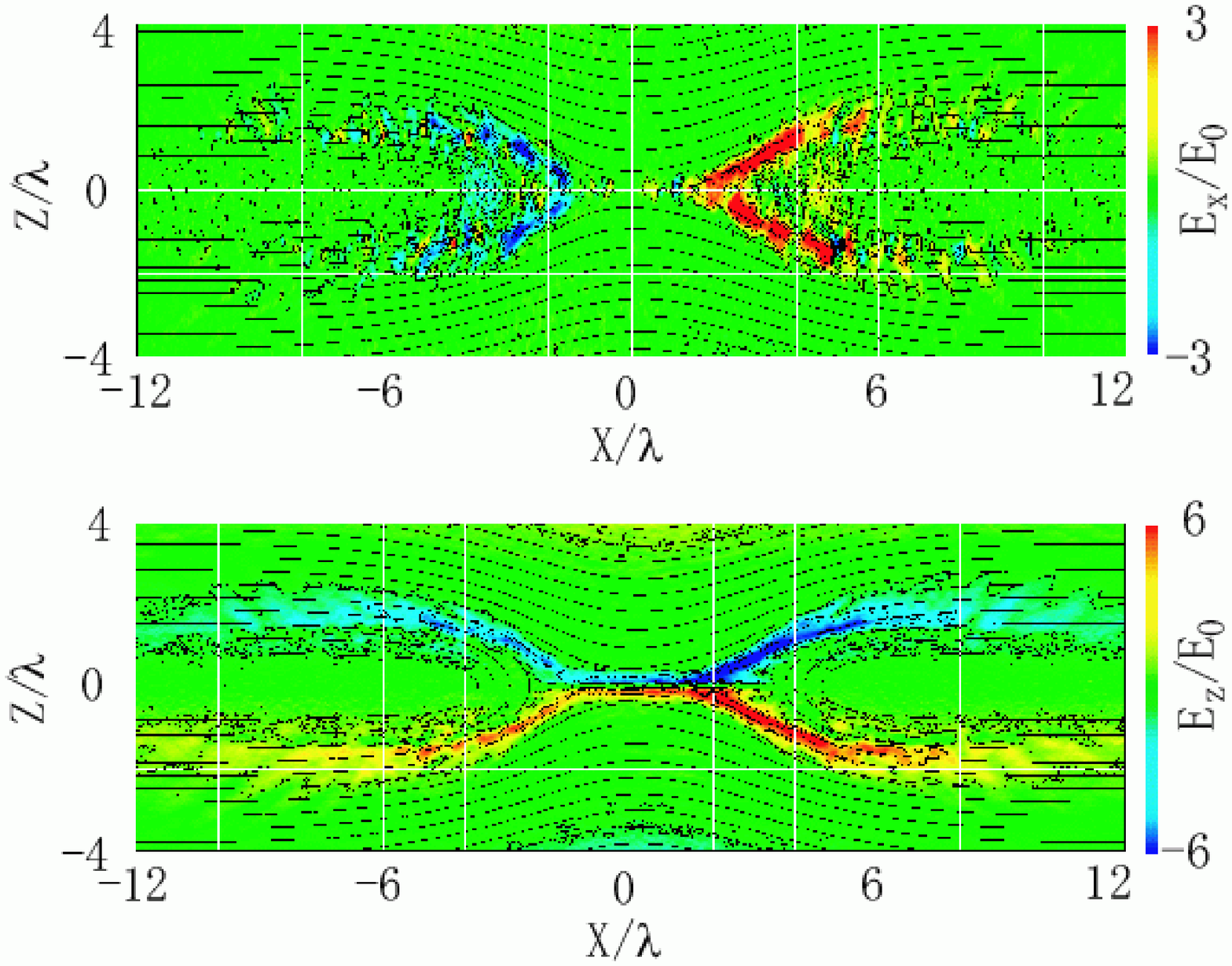}
\caption{The magnetic field lines and the color contours of the electric fields at $t/\tau_A=52.5$ are shown.  Both $E_x$ and $E_z$ are normalized by $E_0=V_A~B_0/c$. Large amplitude polarization electric fields induced around the boundary between the lobe and plasma sheet.  A series of bipolar electric field structure propagating outward from the X-type region along the magnetic field lines can be also seen.  }
\end{figure}

\begin{figure}
\centering
\includegraphics[width=\linewidth]{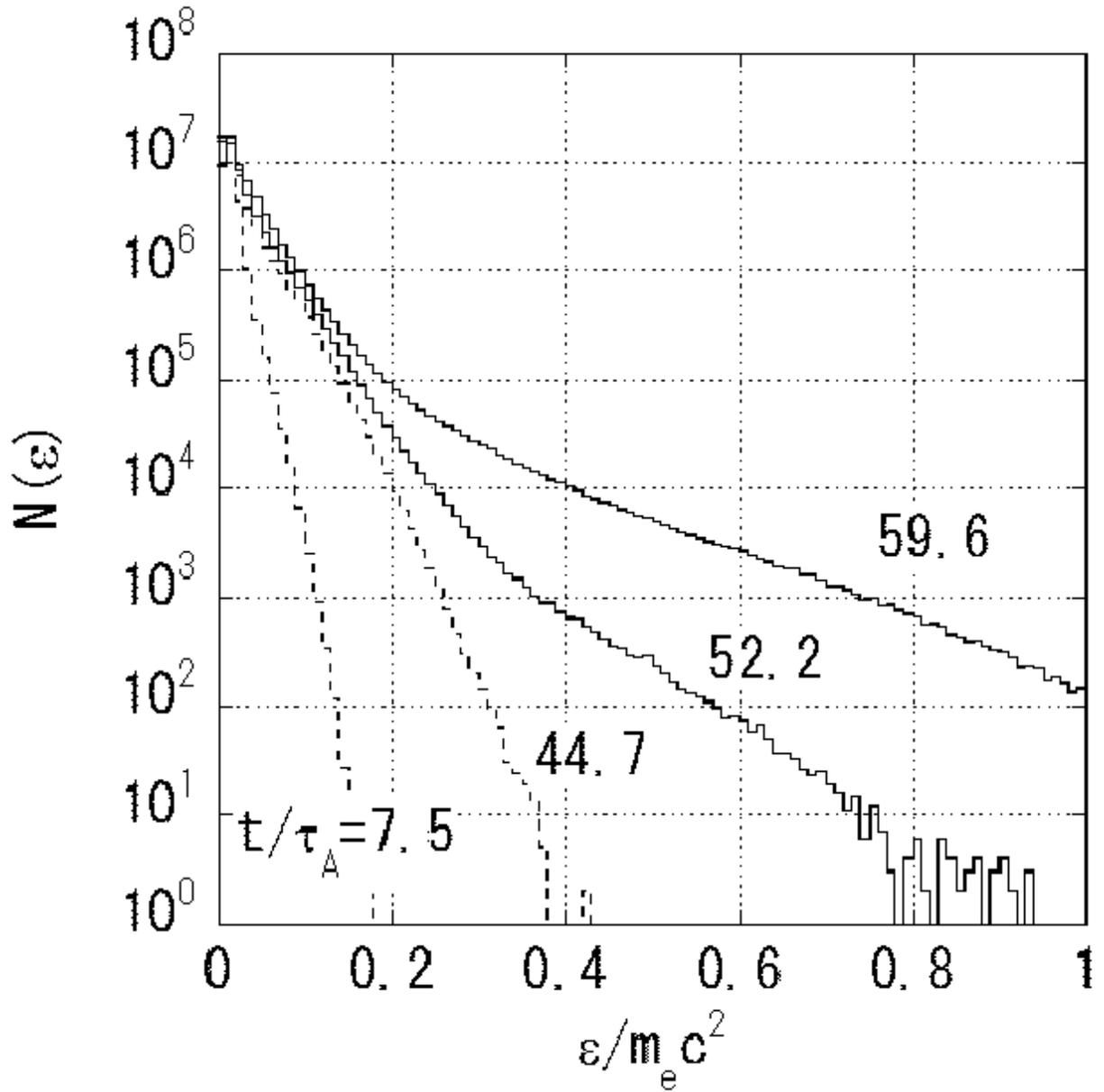}
    \caption{Electron energy spectra integrated over pitch angle in the simulation domain at $t/\tau_A=7.5, 44.7, 52.2$, and $59.6$.}
\end{figure}

\begin{figure}
\centering
\includegraphics[width=\linewidth]{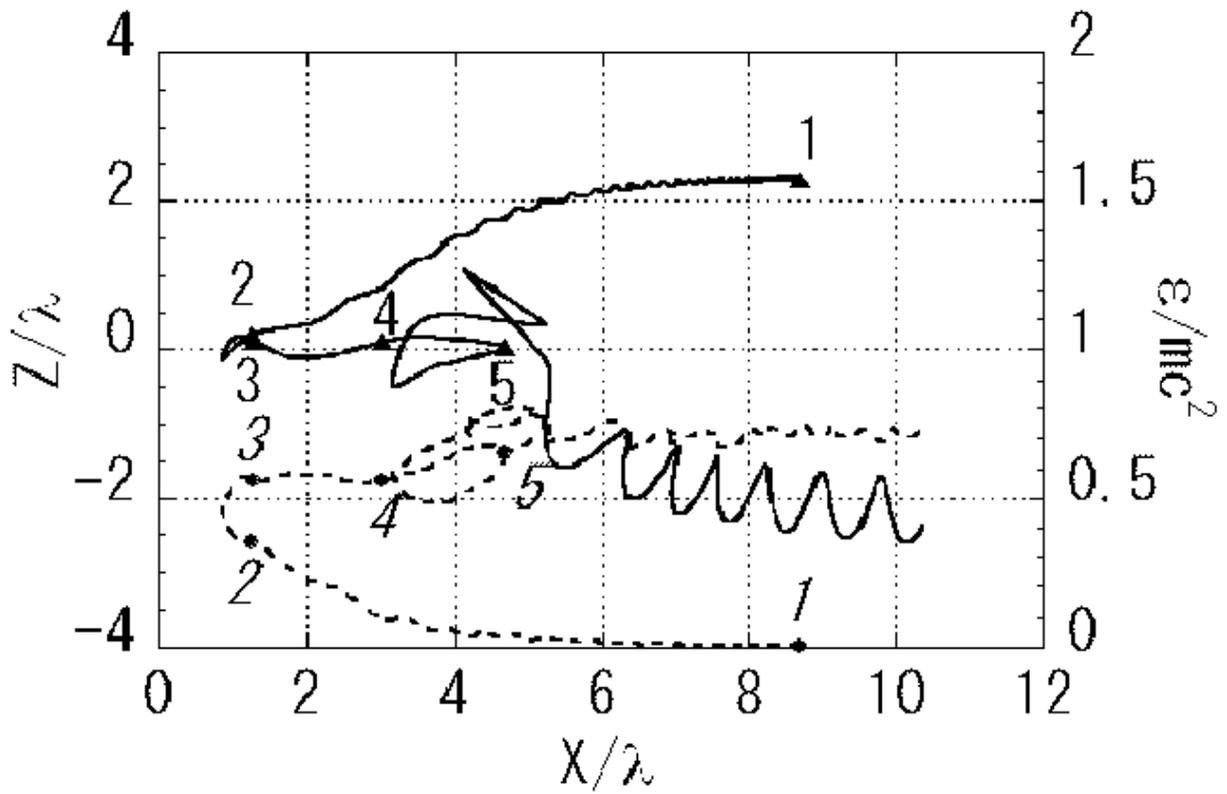}
    \caption{A typical electron trajectory with the surfing acceleration in the $x-z$ plane (solid line), and its energy history as a function of $x$ (dashed line).}
\end{figure}

\begin{figure}
\centering
\includegraphics[width=\linewidth]{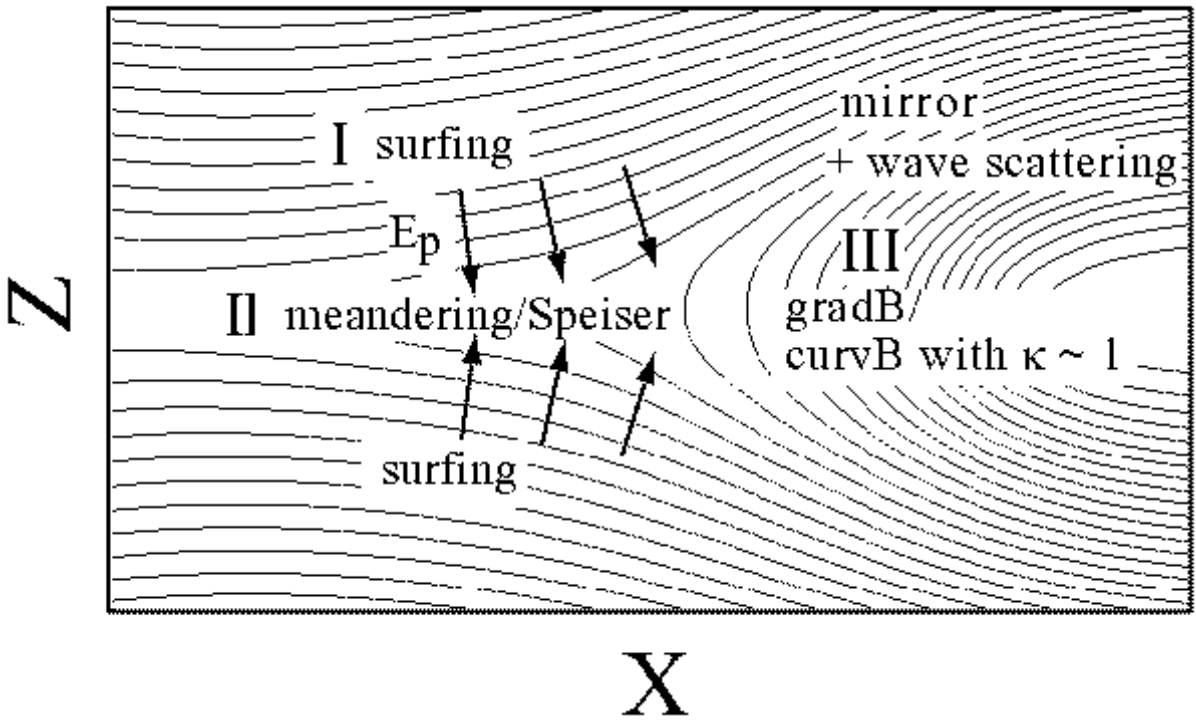}
    \caption{Sketch of acceleration processes during magnetic reconnection.  $E_p$ is the polarization electric field induced in a thin plasma sheet, which plays an important role on surfing acceleration.}
\end{figure}

\begin{figure}
\centering
\includegraphics[width=\linewidth]{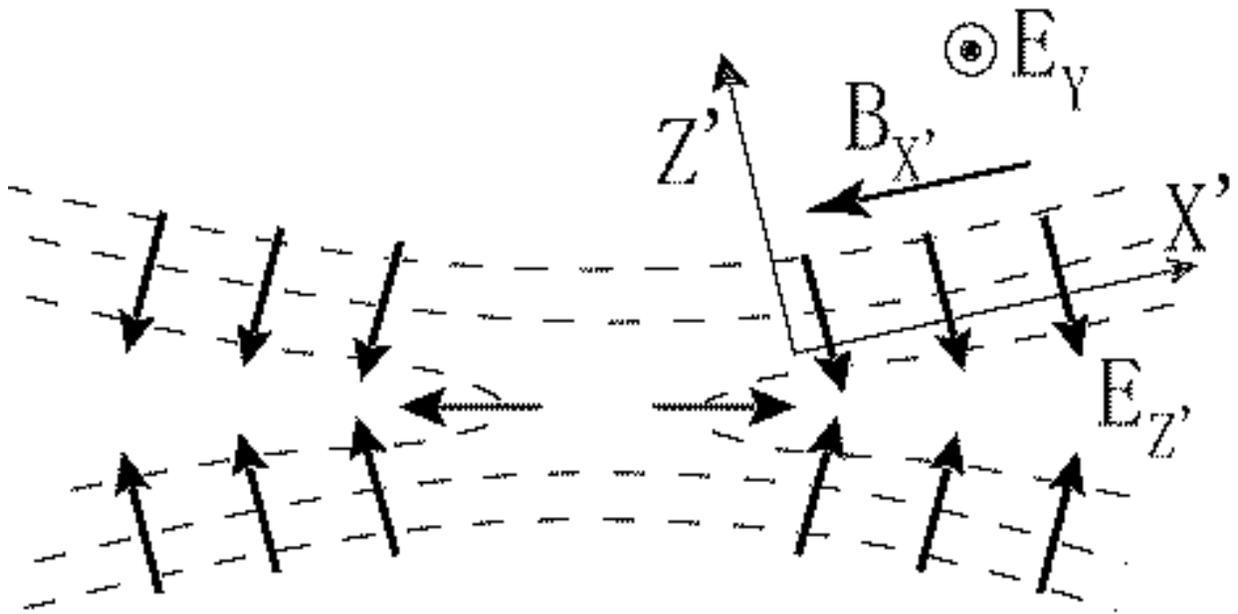}
    \caption{Illustration of the electric and magnetic field structure for electron surfing acceleration in the reconnection region.}
\end{figure}

\begin{figure}
\centering
\includegraphics[width=\linewidth]{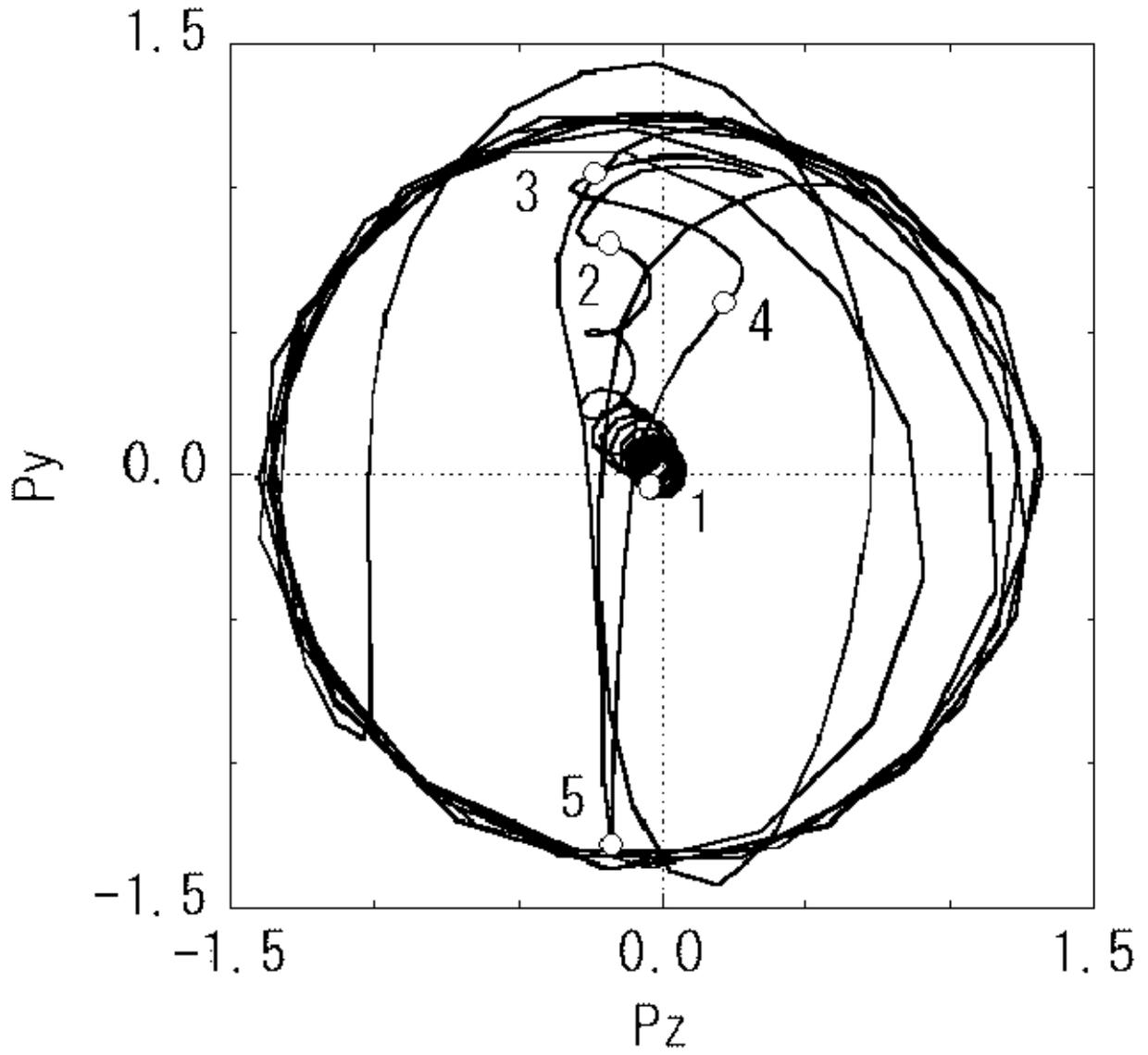}
    \caption{A velocity hodogram for the same electron analyzed in Figure 3.}
\end{figure}

\begin{figure}
\centering
\includegraphics[width=\linewidth]{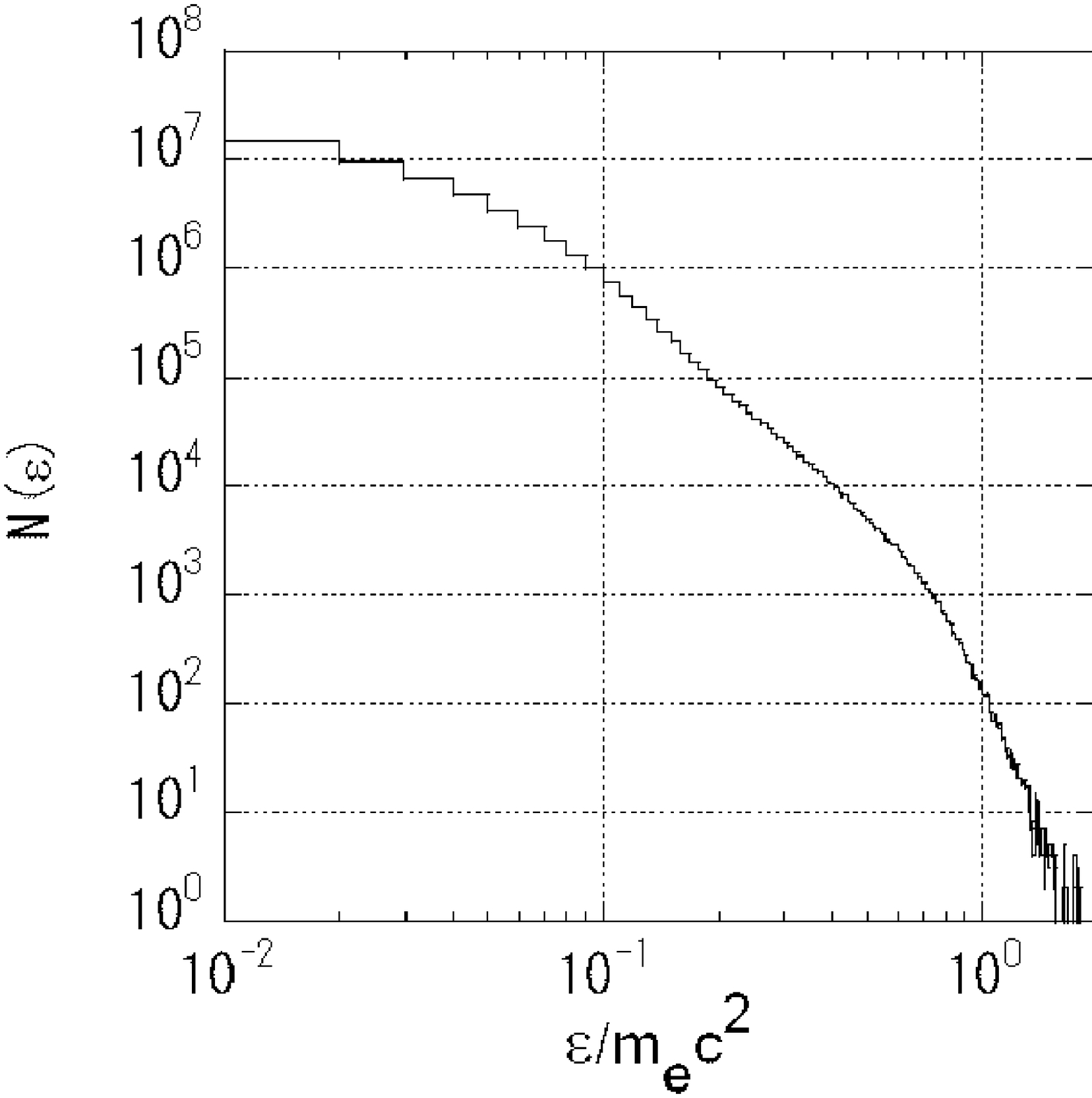}
    \caption{Double power-law energy spectra at $t/\tau_A=59.6$ in the double-logarithmic scale.}
\end{figure}


\begin{thebibliography}{}
\bibitem[]{}
Asano, Y., T. Mukai, M. Hoshino, Y. Saito, H. Hayakawa, and T. Nagai (2003), Evolution of the thin current sheet in a substorm observed by Geotail, \textit{J. Geophys.
Res.}, \textit{108}(A5), 1189,  doi:10.1029/2002JA009785.

\bibitem[]{}
Birn, J. and M. Hesse, (1994), Particle acceleration in the dynamic magnetotail: Orbits in self-consistent three-dimensional MHD fields, \textit{J. Geophys. Res.}, \textit{99}, 109--119.

\bibitem[]{}
Birn, J., J.F. Drake, M.A. Shay, B.N. Rogers, R.E. Denton, M. Hesse, M. Kuznetsova, Z.W. Ma, Z. Bhattacharjee, A. Otto, and P.L. Pritchett, (2001), Geospace environmental modeling (GEM) magnetic reconnection challenge, \textit{J. Geophys. Res.}, \textit{106}(6), 3,715?3,719.

\bibitem[]{}
Birn, J., K. Galsgaard, M. Hesse, M. Hoshino, J. Huba, G. Lapenta, P.L. Pritchett, K. Schindler, L. Yin, J. Buchner, T. Neukirch, and E.R. Priest, (2005), Forced Magnetic Reconnection, \textit{Geophys. Res. Lett.}, \textit{32}, L06105, doi:10.1029/2004GL022058.

\bibitem[]{}
B\"{u}chner, J. and L. M. Zelenyi (1989), Regular and chaotic charged particle motion in magnetotail-like field reversals, 1. Basci theory, \textit{J. Geophys. Res.}, \textit{94}, 11,821--11,842.

\bibitem[]{}
Cattell, C., F. Mozer, K. Tsuruda, H. Hayakawa, M. Nakamura, T. Okada, S. Kokubun, and T. Yamamoto, (1994), Geotail observations of spiky electric fields and low-frequency waves in the plasma sheet and plasma sheet boundary, \textit{Geophys. Res. Lett.}, \textit{21}, 2,987--2,980.

\bibitem[]{}
Christon, S.P., D.J. Williams, D.G. Mitchell, L.A. Frank, and C.Y. Huang (1989), Spectral characteristics of plasma sheet ion and electron populations during undisturbed geomagnetic conditions, \textit{J. Geophys. Res.}, \textit{94}, 13,409--13,424.

\bibitem[]{}
Delcourt, D., J. A. Sauvard, R. F. Martin Jr., and T. E. Moore (1996), On the nonadiabatic precipitation of ions from the near-Earth plasma sheet, \textit{J. Geophys. Res.}, \textit{101}, 17,409--17,418.

\bibitem[]{}
Drake, J. F., N. T. Gladd, and J. D. Huba, (1981), Magnetic field diffusion and dissipation in reversed-field plasmas, \textit{Phys. Fluids}, \textit{24}, 78--87.

\bibitem[]{}
Drake, J., Swisdak, M., Cattell, C., Shay, M. A., Rogers, B. N., and Zeiler, A., (2003), Formation of electron Holes and particle energization during magnetic reconnection, \textit{Science}, \textit{299}, 873--877.

\bibitem[]{}
Fairfield, D. H., (1984), Magneotail energy storage and the variability of the magnetotail current sheet, Magnetic Reconnection in Space and Laboratory Plasmas, AGU/Geophys. Mono., W. Hones, Jr. Ed., 30, 168

\bibitem[]{}
Hayashi, T. and T. Sato, (1978), Magnetic reconnection: Acceleration, heating and shock formation, \textit{J. Geophys. Res.}, \textit{83}, 217--220.

\bibitem[]{}
Hesse, M., M. Kuznetsova, and J. Birn, (2001), Particle-in-cell simulations of three-dimensional collisionless magnetic reconnection, \textit{J. Geophys. Res.}, \textit{106}(6), 29,831--29,841.


\bibitem[]{}
Hoh, F.C., (1966) Stability of sheet pinch, \textit{Phys. Fluids}, \textit{9}, 227.

\bibitem[]{}
Horiuchi, R. and T. Sato (1999) Collisionless magnetic reconnection in the presence of an external driving flow, \textit{J. Plasma Physics}, \textit{61}(3), 415--423.

\bibitem[]{}
Hoshino, M. (1987), The electrostatic effect for the collisionless tearing mode, \textit{J. Geophys. Res.}, \textit{92}(7), 7,368--7,380.

\bibitem[]{}
Hoshino, M. (1991), Forced magnetic reconnection in a plasma sheet with localized resistivity profile excited by lower hybrid drive type instability,\textit{J. Geophys. Res.}, \textit{97}(7), 11,555--11,567.

\bibitem[]{}
Hoshino, M., T. Mukai, T. Yamamoto, and S. Kokubun (1998), Ion dynamics in magnetic reconnection: Comparison between numerical simulation and Geotail observations, \textit{J. Geophys. Res.}, \textit{103}(3), 25,979--25,997.

\bibitem[]{}
Hoshino, M., T. Mukai, T. Terasawa and I. Shinohara (2001a), Suprathermal electron acceleration in magnetic reconnection, \textit{J. Geophys. Res.}, \textit{106}(11),4,509--4,530.

\bibitem[]{}
Hoshino, M., K. Hiraide, and T. Mukai (2001b), Strong electron heating and non-Maxwellian behavior in magnetic reconnection, \textit{Earth Planets Space}, \textit{53}, 627--634.

\bibitem[]{}
Hoshino, M. and N. Shimada (2002), Nonthermal electrons at high Mach number shocks: Surfing Acceleration, \textit{Astrophys. J.}, \textit{572}, 880--887.

\bibitem[]{}
Hoshino, M., (2004), Stratified current sheet during plasma sheet thinning, \textit{COSPAR Colloq. Ser.}, \textit{16}, 108--112.

\bibitem[]{}
Katsouleas, T. and J. M. Dawson, (1983), Unlimited electron acceleration in laser-driven plasma waves, \textit{Phys. Rev. Letters}, \textit{51} (9), 392--395

\bibitem[]{}
Kliem, B., (1994), Particle orbits, trapping, and acceleration in a filamentary current sheet model, \textit{Astrophys. J. Supple. Ser. }, \textit{90}, 719--728.

\bibitem[]{}
Lapenta, G., J. U. Brackbill, (2002), Nonlinear evolution of the lower hybrid drift instability: Current sheet thinning and kinking, \textit{Phys. Plasmas}, \textit{9}(5), 1,544--1,554.

\bibitem[]{}
Lin, R.P., S. Krucker, G.J. Hurford, D.M. Smith, H.S. Hudson, G.D. Holman, R.A. Schwartz, B.R. Dennis, G.H. Share, R.J. Murphy, A.G. Emslie, C. Johns-Krull, N. Vilmer, (2003), RHESSI Observations of particle acceleration and energy release in an intense solar gamma-ray line flare, \textit{Astrophys. J.}, \textit{595} (2), L69--L76,

\bibitem[]{}
McComas, D. J., C. T. Russel, R. C. Elphic, and S. J. Bame, (1986), The near-earth cross-tail current sheet: Detailed ISEE 1 and 2 case studies, \textit{J. Geophys. Res.}, \textit{91}, 4,287--4,301. 

\bibitem[]{}
Mitchell, D. G., D. J. Williams, C. Y. Huang, L. A. Frank, and C. T. Russell, (1990), Current carriers in the near-earth cross-tail sheet during substorm growth phase, \textit{Geophys. Res. Lett.}, \textit{17}, 583--586.

\bibitem[]{}
Mukai, T., M. Hoshino, Y. Saito, I. Shinohara, T. Yamamoto, T. Nagai, and S. Kokubun, (1998), Pre-onset and onset signatures for substroms in the near-tail plasma sheet: GEOTAIL observations, Substorms-4, S. Kokubun and Y. Kamide, Eds., Terra Sci. Publ., 131--136 

\bibitem[]{}
Nagai, T., I. Shinohara, M. Fujimoto, M. Hoshino, Y. Saito, S. Machida, and T. Mukai (2001), Geotail observations of the Hall current system: Evidence of magnetic reconnection in the magnetotail, \textit{J. Geophys. Res.}, \textit{106}(11) 25,929--25,959.

\bibitem[]{}
Nishida, A., (2000), The Earth's dynamic magnetotail, \textit{Spaace Sci. Rev.}, \textit{91}, 507--577.

\bibitem[]{}
Pritchett, P. L., and F. V. Coroniti (1995), Formation of thin current sheets during plasma sheet convection, \textit{J. Geophs. Res.}, \textit{100}, 23,551--23,565. 

\bibitem[]{}
Pritchett, P.L., (2001), Collisionless magnetic reconnection in a three-dimensional open system, \textit{J. Geophs. Res.}, \textit{106} (11), 25,961--25,5977.


\bibitem[]{}
Pulkkinen, T. I., D. N. Baker, D. G. Mitchell, R. L. McPherron, C. Y. Huang, and L. A. Frank, (1994), Thin current sheets in the magnetotail during substorms: CDAW 6 revisited, \textit{J. Geophys. Res.}, \textit{99}, 5,793--5,803. 

\bibitem[]{}
Sagdeev, R.A. and V.D. Shapiro, (1973), Influence of transverse magnetic-field on Landau damping, \textit{JETP Lett. Engl. Transl.}, \textit{17}(7), 279--282.

\bibitem[]{}
Sato, T., H. Matsumoto, and K. Nagai, (1982), Particle acceleration in time-developing magnetic reconnection process, \textit{J. Geophys. Res.}, \textit{87}, 6,089--6,097.

\bibitem[]{}
Scholer, M. and F. Jamitzky, (1987), Particle orbits during the development of plasmoids, \textit{J. Geophys. Res.}, \textit{92}, 12,181--12,186.

\bibitem[]{}
Scholer, M., I. Sidorenko, C.H. Jaroschek, R.A. Treumann, and A. Zeiler, (2003), Onset of collisionless magnetic reconnection in thin current sheets: Three-dimensional particle simulations, \textit{Phys. Plasmas}, \textit{10}(9), 3,521--3,527.

\bibitem[]{}
Sergeev, V. A., D. G. Mitchell, C. T. Russell, and D. J. Williams, (1993), Structure of the tail plasma/current sheet at ~11Re and its changes in the course of a substorm, \textit{J. Geophys. Res.}, \textit{98}, 17,345--17,365. 

\bibitem[]{}
Shinohara, I., H. Suzuki, M. Fujimoto, M. Hoshino, (2001), Rapid large-scale magnetic-field dissipation in a collisionless current sheet via coupling between Kelvin-Helmholtz and lower-hybrid-drift instabilities,  \textit{Phys. Rev. Letts.}, \textit{87-9}, 5001.

\bibitem[]{}
Speiser, T. W. Particle trajectories in model current sheet 1, Analytical solution (1965), \textit{J. Geophys. Res.}, \textit{70}, 4,219--4,226.

\bibitem[]{}
Zelenyi, L.M., H.V. Malova, V.Yu. Popov, D. Delcourt, and A.S. Sharma (2004), Nonlinear equilibrium structure of thin current sheets: influence of electron pressure anisotropy, \textit{Nonlinear Proc. Geophys.}, \textit{11}, 579--587.

\end{thebibliography}
\end{document}